\documentclass[%
reprint,
superscriptaddress,
 amsmath,amssymb,
 aps, prb,
]{revtex4-2}

\usepackage[dvipsnames]{xcolor}
\usepackage[colorlinks]{hyperref}
\hypersetup{
colorlinks=True,
linkbordercolor=White,
linkcolor=BrickRed,
citecolor=Blue,	 
}

\usepackage{siunitx}
\usepackage{graphicx}
\usepackage{layouts}
\usepackage[T1]{fontenc}

\newcommand\YMN[1]{\textcolor{black}{#1}}

\renewcommand\vec[1]{\mathbf{#1}}

\begin{document}
\vspace{-0.5cm}

\title{
Valley Splittings in Si/SiGe Heterostructures from First Principles}
%

\author{Lukas Cvitkovich}
\email[]{cvitkovich@iue.tuwien.ac.at}
\affiliation{Institute for Microelectronics, Technische Universität Wien}


\author{Tancredi Salamone}
\affiliation{Univ. Grenoble Alpes, CEA, LETI, F-38000, Grenoble, France}

\author{Christoph Wilhelmer}
\affiliation{Institute for Microelectronics, Technische Universität Wien}

\author{Biel Martinez}
\affiliation{Univ. Grenoble Alpes, CEA, LETI, F-38000, Grenoble, France}

\author{Tibor Grasser}
\affiliation{Institute for Microelectronics, Technische Universität Wien}

\author{Yann-Michel Niquet}
\affiliation{Univ. Grenoble Alpes, CEA, IRIG-MEM-L Sim, F-38000, Grenoble, France}


\begin{abstract}
We compute valley splittings in Si/SiGe superlattices using \textit{ab initio} density functional theory (DFT). This first-principle approach is expected to provide an excellent description of interfaces, strains, and atomistic disorder without empirically fitted parameters. \YMN{We benchmark atomistic tight-binding (TB) and the ``$2k_0$'' theory within the effective mass (EM) approximation against DFT. We show that DFT supports the main conclusions of the $2k_0$ theory, but reveals some limitations of semi-empirical methods such as the EM and TB, in particular about the description of atomistic disorder. The DFT calculations also highlight the effects of strong valley-orbit mixing at large valley splittings. Nevertheless, TB and the $2k_0$ theory shall provide reasonable valley splitting statistics in many heterostructures of interest for spin qubit devices.}
\end{abstract}

\maketitle

\section{Introduction}

Quantum dots in Si/SiGe heterostructures are among the most promising candidates for the implementation of large-scale arrays of spin qubits for quantum computing~\cite{Loss1998,burkard_semiconductor_2023,Zwanenburg2013}. Epitaxial Si/SiGe interfaces are, indeed, of much better quality than silicon/oxide interfaces, and the disordered gate stack can be moved tens of nanometers away from the active Si layer. Moreover, silicon, as well as germanium, can be isotopically purified~\cite{itoh_isotope_2014, Struck2020, cvitkovich_hf_2024} to eliminate the nuclear spins whose hyperfine interactions with the electron spins can limit coherence. Coherent~\cite{Struck2020,yoneda_quantum-dot_2017,mills_two-qubit_2022,weinstein_universal_2023} and high-fidelity~\cite{takeda_resonantly_2020,blumoff_fast_2022,philips_universal_2022,mills_high-fidelity_2022} spin qubits have been demonstrated in Si/SiGe heterostructures, with the best fidelities of one and two-qubit gates exceeding the threshold for quantum error correction~\cite{Noiri2022,Xue2022}.

One of the main challenges for electron spin qubits in silicon is the existence of nearly degenerate states near the conduction band edge. Indeed, silicon is an indirect band gap material with six equivalent conduction band valleys near the $\pm X$, $\pm Y$ and $\pm Z$ points of the first Brillouin zone~\cite{Zwanenburg2013}. Although the degeneracy between the $X$, $Y$ and $Z$ valleys is lifted by vertical confinement in the heterostructure~\cite{Boykin04}, the splitting $E_\mathrm{VS}$ between the ground-state $Z$ valleys remains usually small and may be comparable to the Zeeman splitting between the spin states. This provides a leakage channel for the quantum information encoded in the spin that is detrimental to the coherence of the qubits~\cite{Zwanenburg2013}. Moreover, current devices suffer from the large variability of the valley splittings, which can range from a few tens to a few hundreds of $\mu$eV~\cite{Shi2011,Kawakami2014,Mi2017,Oh2021} even on the same wafer. Similarly, recent efforts to establish mid-range links between different quantum dots (by ``shuttling'' electrons) may be compromised by spatially varying valley splittings~\cite{Langrock2023,Volmer2024,Losert_shuttling_2024}. Experimental evidence supports theoretical assumptions that these variations result from atomic fluctuations at the interfaces and in the SiGe alloys~\cite{PhysRevB.82.205315,Neyens2018,Dodson2022,Wuetz2022,Lima2023,Esposti2024,Pena2024,Lima2024, Klos2024}.

Several innovative quantum well designs have been put forward to enhance the valley splittings, such as the introduction of Ge spikes~\cite{McJunkin2021} or oscillating Ge concentrations (so-called wiggle wells)~\cite{McJunkin2022,Feng2022}. Such designs boost the inter-valley scattering by introducing short wavelength components in the confinement potential able to couple the opposite $Z$ valleys. This strategy is backed by the ``$2k_0$'' theory within the effective mass (EM) approximation, which shows that $E_\mathrm{VS}$ is proportional to the Fourier component of the confinement potential at wave number $q_z=2k_0$, where $2k_0=1.7\times 2\pi/a$ is the reciprocal distance between the $\pm Z$ valleys and $a$ the lattice parameter of silicon~\cite{Losert2023}.

The $2k_0$ theory is, however, a semi-empirical model that makes potentially severe approximations about the description of alloy disorder and about the details of the potential around the fast variations of the Ge composition that give rise to inter-valley scattering. It assumes, in particular, that this potential only depends on the {\it local} Ge concentration, thus neglecting the effects of charge transfers and other non-local corrections resulting from the specific arrangement of atoms in the alloy. \YMN{Atomistic tight-binding (TB) methods~\cite{Boykin04,Niquet2009} are expected to provide a better description of alloy disorder but still miss physics beyond nearest neighbor interactions.} In order to test the validity of these assumptions, we use \textit{ab-initio} density functional theory (DFT) as a modeling benchmark for the valley splittings in Si/Si$_{0.7}$Ge$_{0.3}$ superlattices. DFT seamlessly integrates the complex interplay between quantum well design, atomistic disorder, local charge transfers, and strains from first principles~\cite{Cvitkovich_SISPAD2024}, and does not rely on empirically fitted parameters like the EM~\cite{Friesen2007,PhysRevB.82.205315,Lima2023} or tight-binding (TB) methods~\cite{Boykin04,Kharche2007,Boross2016}. We compare the DFT valley splittings with TB data on the same atomic structures. Our results support the main trends and predictions of the $2k_0$ theory while also revealing some shortcomings of the EM and TB methods. \YMN{In particular, the EM (and, to a lesser extent, TB) miss some effects of atomistic disorder. Moreover, DFT shows that valley-orbit mixing (beyond the $2k_0$ theory) can be very significant when the valley splittings are large.}

The paper is organized as follows: we first introduce the heterostructures and modeling framework in Section \ref{sec:methodology}, then discuss valley splittings in Si/Si$_{0.7}$Ge$_{0.3}$ quantum wells with smooth interfaces and in wiggle wells in Section \ref{sec:results}.

\section{Methodology}
\label{sec:methodology}

\subsection{Heterostructure and supercell}

\begin{figure}[htbp]
    \centerline{\includegraphics[width=\linewidth]{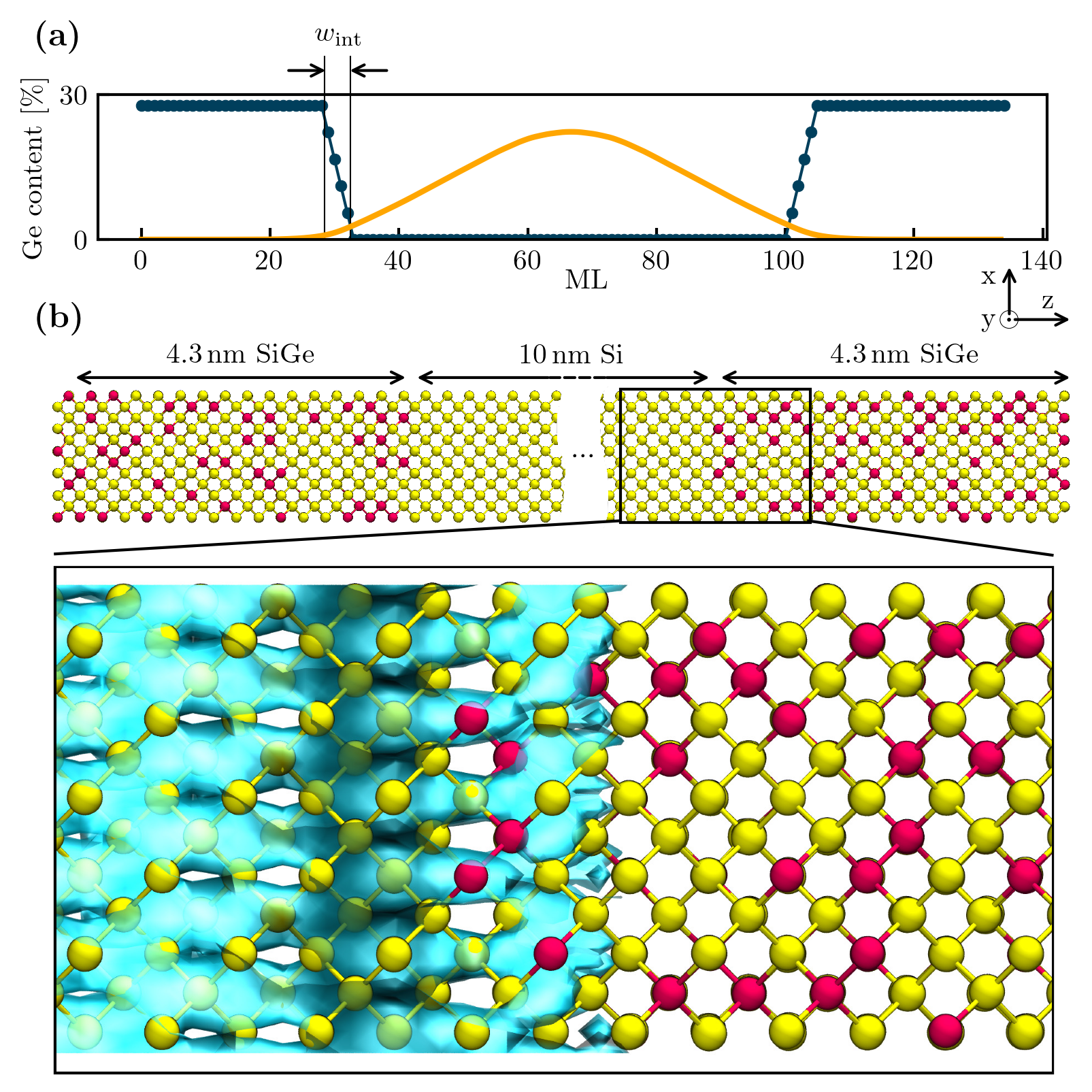}}
    \caption{
        \textbf{(a)}~\YMN{Ge concentration profile (blue circles) for smooth Si/SiGe interfaces with width $w_\mathrm{int}=4$\,MLs in a supercell with size $n=3$. At the interfaces, the Ge concentration varies from 0 to 27.8\% by steps of $\approx5.56\%$ ($w_\mathrm{int}$ corresponds to the number of MLs with intermediate concentrations). The orange line shows the envelope of the DFT ground-state conduction band wave function at zero electric field.}
        \textbf{(b)}~\YMN{Atomistic 3D model (side view) of a SiGe/Si/SiGe heterostructure with \YMN{abrupt interfaces ($w_\mathrm{int}=0$)}. The supercell (with size $n=3$) is 1.649$\times$1.649$\times$18.5$\,$nm$^3$ and contains 2448 atoms (Si in yellow, Ge in red). The bottom panel shows an iso-density surface (blue) of the DFT wave function, highlighting the penetration in the disordered SiGe alloy.}
    }
    \label{fig:model}
\end{figure}

We compute the splitting between $Z$ valleys in planar [001] heterostructures at in-plane wave vector $k_\perp=0$ (no lateral confinement). For that purpose, we consider a superlattice with $\approx 10$\,nm (72 monolayers) thick Si quantum wells separated by $\approx 8.6$\,nm (64 monolayers) thick Si$_{0.7}$Ge$_{0.3}$ barriers (see Fig.~\ref{fig:model}). The Ge concentration in the barriers is typical of recent experiments~\cite{philips_universal_2022,mills_two-qubit_2022,mills_high-fidelity_2022,weinstein_universal_2023}. \YMN{These heterostructures are modeled as supercells with in-plane area $S=na\times na$ (with $a$ the in-plane lattice parameter and $n$ a positive integer) and periodic boundary conditions in all directions.}

We analyze various model structures, where Ge atoms are randomly distributed on the lattice sites according to a predefined Ge concentration profile (illustrative profile shown in Fig.~\ref{fig:model}a). \YMN{The total number of Ge atoms in each monolayer (ML), $n_\mathrm{Ge}=E(2xn^2)$, is set by the Ge concentration $x$ in that ML and is independent of the random realization of the supercell [$E(u)$ is the integer nearest to $u$; this assumption will be further discussed in Section \ref{subsec:smoothed}]}. Our models include sharp interfaces, smooth interfaces with a width of several MLs, and wiggle wells (with oscillating Ge concentration)~\cite{McJunkin2022,Feng2022}. \YMN{The in-plane lattice parameter $a$ is matched to a virtual Si$_{0.7}$Ge$_{0.3}$ buffer ($a=5.5216$\AA) so that a pure Si layer ($a_0=5.4588$\AA\ within our DFT framework) experiences a tensile, biaxial strain $\varepsilon_\parallel=1.15\%$}. The atomic positions within the supercell and its length along the growth axis $z$ are relaxed in order to adapt to the Ge content. In addition to the imposed in-plane strain, the model thus features local lattice distortions in all three dimensions due to the alloy disorder. We sample different random realizations of the same Ge concentration profile in order to assess the effects of this disorder.

\subsection{DFT and TB calculations}
\label{sec:DFTTB}

DFT as implemented in the CP2K code~\cite{CP2K} is used to optimize the geometry of the superlattices and compute their electronic structure. CP2K relies on atom-centered basis sets, which allow for the efficient simulation of large systems. \YMN{The structures are relaxed using the PBE exchange-correlation functional~\cite{PBE} until the maximum residual forces on the atoms are $<3\times 10^{-5}$\,Ha/Bohr and the maximum displacement of the atoms between two successive optimization steps is $<5\times10^{-5}$\,nm. The valley splittings are then computed in these relaxed structures with the hybrid functional PBE0~\cite{PBE0_TC_LRC}}, which is known to give accurate band gaps in semiconductor materials~\cite{Functionals2011}. It includes a fraction of the exact Hartree-Fock exchange. The orbitals are expanded in double-$\zeta$ basis sets and the core electrons are approximated by Goedecker-Teter-Hutter (GTH) pseudopotentials~\cite{GTH}. Static vertical electric fields $F_z$ are applied using the formalism of Ref.~\cite{Umari2002} (see Appendix~\ref{app:field} for details). 

The TB calculations are performed with the first nearest-neighbor $sp^3d^5s^*$ model of Ref.~\cite{Niquet2009} that accounts for the effects of strains. We use the DFT atomic positions as input for TB \footnote{The equilibrium Si-Si, Ge-Ge and Si-Ge bond lengths of the TB model are thus matched to the DFT bond lengths to avoid the introduction of spurious hydrostatic strains in the TB description.} to enable unambiguous comparisons between the two approaches. However, DFT is not applicable to the million-atom systems TB models are originally intended for. The structures used as input for TB calculations are thus usually relaxed with semi-empirical models such as Keating's valence force field (VFF)~\cite{Keating66,Niquet2009}. Therefore, we also compare in this work the TB valley splittings computed with DFT and with VFF geometries.

In both DFT and TB, the valley splitting $E_\mathrm{VS}$ is calculated as the difference between the lowest two sub-bands at $k_\perp=0$. We do not take spin-orbit coupling into account.

One of the main constraints of DFT is computational cost, which limits the side $n$ of the supercell. \YMN{In the present work, we set $n=3$ in order to allow one-to-one comparisons between DFT and TB}. This implies that the Ge fraction in each ML is discretized and can only vary in steps of $1/18\approx 5.56\%$ (see Fig.~\ref{fig:model}). The actual Ge concentration in the buffer and barrier is thus 27.8\%. \YMN{We discuss in Appendix~\ref{app:scaling} the scaling of the valley splitting statistics with respect to the supercell side.}

\subsection{Properties and interpretation of the DFT model}
\label{sec:interpretation}

Although many concepts of the EM theory are transferable to DFT, the language of the two approaches is not fully compatible. In order to bridge this gap, we review the properties and interpretation of our \textit{ab-initio} framework, and highlight the differences between DFT and the EM/TB.

According to Refs.~\cite{Wuetz2022,Feng2022,Losert2023}, the valley splitting can be expressed within the EM theory as
\begin{equation}
\label{eq:matrix_element}
E_\mathrm{VS}\approx2|\Delta|\text{ with }\Delta=\langle \Psi_{+z} | V_\mathrm{qw} | \Psi_{-z} \rangle\,,
\end{equation}
where $\Psi_{\pm z}(z)=\mathrm{e}^{\pm i k_0 z} \Phi(z)$ are the Bloch waves of the $\pm Z$ valleys at $k_z=\pm k_0=\pm 0.85\times 2\pi/a_0$, modulated by the envelope function $\Phi(z)$ resulting from confinement in the quantum well potential $V_\mathrm{qw}(z)$ (the conduction band offset between the strained SiGe alloy at $z$ and the buffer). Namely, the valley splitting is proportional to the modulus of the $q=2k_0$ component of the Fourier transform of $V_\mathrm{qw}(z)|\Phi(z)|^2$. As a first-order expression, Eq.~\eqref{eq:matrix_element} assumes that the envelope $\Phi(z)$ is the same for both valley states (no valley-orbit hybridization~\cite{PhysRevB.88.035310}). The coupled valley eigenstates read:
\begin{equation}
\label{eq:Psi12}
\tilde\Psi_{1,2}(\vec{r})=\frac{1}{\sqrt{2}}\left[\Psi_{+z}(z)u_{+z}(\vec{r})\pm \Psi_{-z}(z)u_{-z}(\vec{r})\right]\,,
\end{equation}
where $u_{\pm z}(\vec{r})$ are the lattice-periodic Bloch functions of the $\pm Z$ valleys. \YMN{Although validated by extensive comparisons with TB calculations~\cite{Wuetz2022,Losert2023}, this ``$2k_0$'' theory is not free from approximations (see in particular the discussion about inter-valley potentials in Ref.~\cite{Salamone25}). Also, the EM theory (and, to a lesser extent, TB) is not expected to hold for rapidly varying components of the potential with wave numbers as large as $2k_0$.} It is, therefore, relevant to benchmark Eq.~\eqref{eq:matrix_element} against DFT calculations, which best describe the effects of alloy disorder and charge transfers that may reshape the potential, especially near the interfaces.

Moreover, the confinement potential $V_\mathrm{qw}$ and the envelope $\Phi$ are rather ill-defined concepts from the point of view of DFT. In particular, DFT provides {\it full} wave functions including the atomistic Bloch function modulations washed out in envelope theories. The DFT wave functions could, therefore, at best be mapped onto Eq.~\eqref{eq:Psi12}, but the Bloch functions $u_{\pm z}(\vec{r})$ are actually ill-defined in a random alloy~\cite{Abrikosov98}. There are, therefore, inherent ambiguities in the definition of envelope functions in DFT (as well as in TB) calculations, which reflect the assumptions of the EM theory.

\begin{figure}[htbp]
    \centerline{\includegraphics[width=\linewidth]{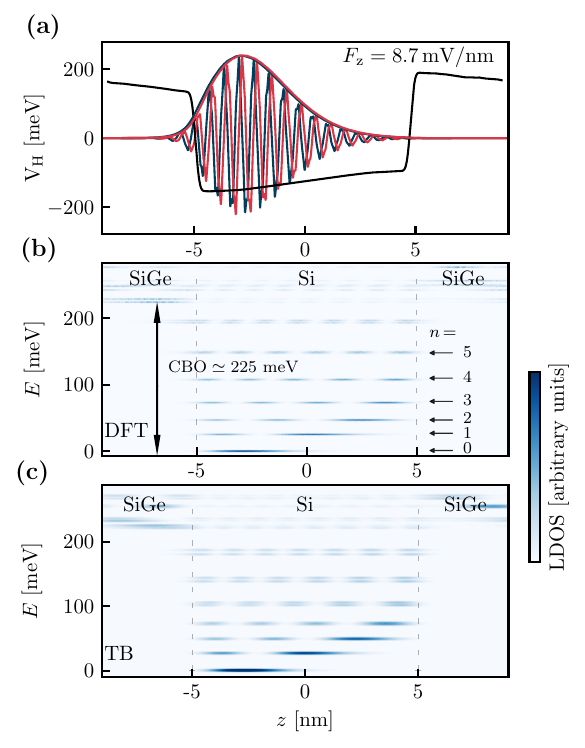}}
    \caption{
        \textbf{(a)}~Cross-section averaged ground-state valley wave functions $\tilde\Psi_{1,2}(z)$ (blue, red) confined in a quantum well with sharp interfaces characterized by the Hartree potential $V_\mathrm{H}(z)$ (black). The electron is squeezed on the bottom interface by an external electric field $F_\mathrm{z}=8.7$\,mV/nm. The wave functions are real and have similar envelopes $\tilde\Phi_{1,2}(z)$ but are phase shifted with respect to each other.
        \YMN{\textbf{(b)}~DFT LDOS in the same quantum well as in (a). The orbital sub-bands with $n=0,1,\dots$ nodes can be identified within the Si layer. Valley splittings in the range of 0.1\,meV are too small to be clearly resolved on this energy scale. \textbf{(c)}~TB LDOS in the same quantum well as in (a) and (b). The external potential extracted from the DFT calculation in (a) has been transferred to the TB Hamiltonian as explained in Appendix \ref{app:field}. The DFT and TB LDOS are in excellent agreement.}
    }
    \label{fig:DFTsolutions}
\end{figure}

As an illustration, the lowest two conduction band wave functions $\tilde\Psi_{1,2}(\vec{r})$ of a representative heterostructure are shown in Fig.~\ref{fig:DFTsolutions}a. The Si/SiGe interfaces of this heterostructure are smoothed over 4\,ML (see profile in Fig.~\ref{fig:model}a), and a vertical electric field $F_z=8.7$\,meV/nm is applied to confine the electron on the bottom interface and suppress the interferences with the top one. Since the properties of a planar heterostructure (without lateral confinement) can be described by in-plane averaged quantities, we use cross-section averaged wave functions and potentials for our analysis. As expected, the coupled valley wave functions show oscillations with period $k_0\approx 0.85\times 2\pi/a_0$ and are phase shifted with respect to each other. We build an estimate for $\Phi^2(z)$ by convoluting the squared DFT wave functions with a Gaussian (with standard deviation $\sigma=0.25$\,nm). This is essentially a low-pass filter that smooths out short-wavelength oscillations and returns the slowly varying envelope \footnote{\YMN{We build envelopes in the same way in TB by summing Gaussians weighted by the total probability of presence of each atom (the sum of the squared modulus of the orbital coefficients).}}.

The Si/SiGe quantum well potential $V_\mathrm{qw}$ is also not uniquely defined in DFT. In particular, the exchange-correlation potential is non-local in the present hybrid DFT calculations, which include a fraction of the exact Hartree-Fock exchange that can not be represented by a local operator. Although this may be considered as a diversion of hybrid functionals (the exact Kohn-Sham potential being by design local), non-locality is in fact intrinsic to all many-body post-DFT methods such as the \textit{GW} approximation~\cite{GW,GW_2}. Non-local effects may be more significant in highly inhomogeneous systems, and in particular near sharp variations of the Ge concentration that make large contributions to the $2k_0$ theory.

The only non-ambiguous local component of the many-body Hamiltonian is the Hartree potential $V_\mathrm{H}$, which is the total electrostatic potential of ions and electrons but does not contain any information about exchange and correlation. This potential is plotted in Fig.~\ref{fig:DFTsolutions}a; it is also averaged in the cross-section of the heterostructure and low-pass filtered by convolution with a Gaussian (same procedure as for the envelope $\Phi$). In a first approximation, we can estimate the conduction band offset between Si and Si$_{0.7}$Ge$_{0.3}$ from the steps of $V_\mathrm{H}(z)$ at the interfaces. This estimate, $\mathrm{CBO}=280$\,meV, is significantly larger than the value $\mathrm{CBO}=170$\,meV generally used in EM theories~\cite{Maiti1998}. However, it misses contributions from the exchange and correlation.

We can make a refined assessment of the CBO on the local density of states (LDOS) of the superlattice, plotted in Fig.~\ref{fig:DFTsolutions}b. The LDOS is basically a representation of the squared envelopes (as defined above) as a function of position and energy. It provides a comprehensive picture of the electronic structure of the quantum well. The orbital sub-bands, that are separated by a few tens of meV, can be characterized by the number of spatial nodes $n=0,1,\dots$. Each peak (along the energy axis) is actually a doublet split by the valley mixing, although the gap is hardly visible at the scale of Fig.~\ref{fig:DFTsolutions}b, except for highly excited states. The separation between the ground and the first unbound orbitals at $F_z=0$ gives a much better estimate of the CBO, as confirmed by simple effective mass calculations. The net CBO, including (non-local) exchange and correlation effects, is therefore closer to $225$\,meV, in slightly better agreement (though still $\approx30\%$ larger) with the experimental value $\mathrm{CBO}=170$\,meV~\cite{Maiti1998}.

\YMN{The energy and shape of the wave functions (in particular, the penetration into the SiGe barriers) are expected to be dependent on the CBO. Therefore, in order to make consistent comparisons between DFT and TB, we have adjusted the orbital energies of the TB model \footnote{\YMN{Namely, we have shifted the orbital energies of the Ge atoms so that the unstrained valence band offset between Si and Ge is $\mathrm{VBO}=0.91$\,eV~\cite{Niquet2009}.}} to reproduce the DFT $\mathrm{CBO}=225$\,meV. With this correction, the TB LDOS is in excellent agreement with the DFT LDOS, as illustrated in Fig.~\ref{fig:DFTsolutions}c. The DFT and TB confinement masses are, in particular, close enough so that the two methods yield comparable sub-band splittings.}

\section{Results}
\label{sec:results}

In this section, we discuss the valley splitting $E_\mathrm{VS}$ in smoothed and wiggle wells, and compare DFT with TB calculations. \YMN{The same superlattices (same distribution of Ge atoms and atomic positions) are therefore computed with both DFT and TB. As discussed above, the original CBO of the TB model (150\,meV) has been tuned to match the CBO of DFT (225\,meV), in order to focus on the intrinsic differences between the two approaches. Also, the same vertical electric field profiles are used in TB and DFT. For that purpose, the external potential extracted from the DFT calculations is transferred to the TB Hamiltonian as explained in Appendix \ref{app:field}. The vertical electric fields $F_z$ quoted in this work correspond to the slope of the Hartree potential in the Si well.}

\subsection{Smoothed Si quantum wells}
\label{subsec:smoothed}

\begin{figure}[tbp]
    \centerline{\includegraphics[width=\linewidth]{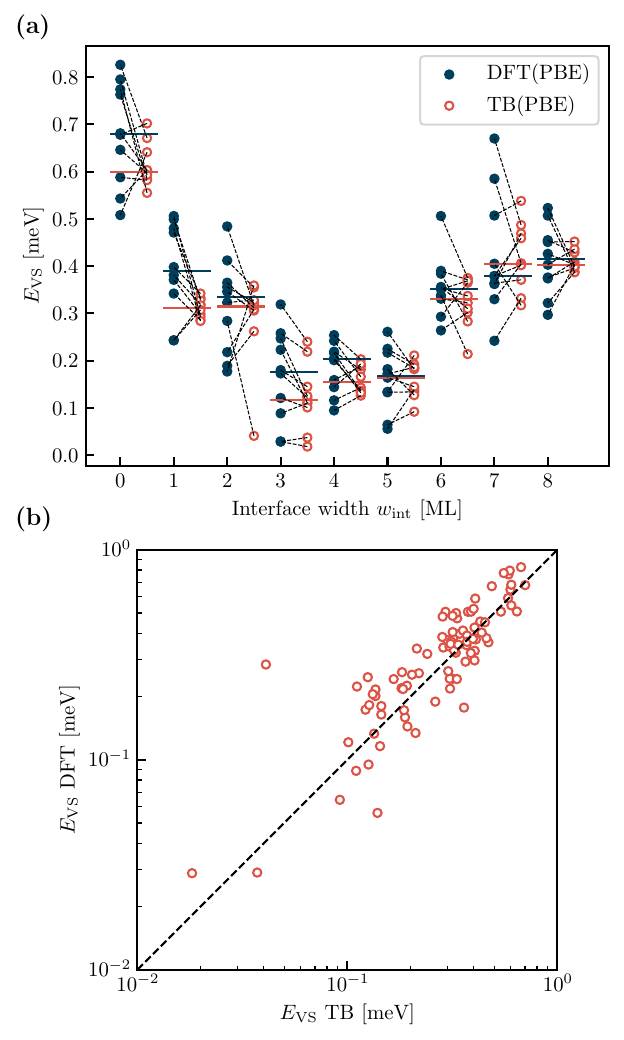}}
    \caption{
        \textbf{(a)} DFT and TB valley splittings $E_\mathrm{VS}$ calculated in Si wells with smooth interfaces characterized by the width $w_\mathrm{int}$ (see text). \YMN{The atomic positions relaxed with the PBE exchange-correlation functional are used as input for DFT and TB calculations.} The plot shows the distribution of ten random realizations of each $w_\mathrm{int}$, with the DFT and TB valley splittings of the same structures connected by dashed lines. The median value of each distribution is indicated by a horizontal bar. 
        \textbf{(b)} Correlation between the DFT and TB valley splittings in Si wells with smooth interfaces (all $w_\mathrm{int}$'s and samples). The dashed line is a guide to the eye with unity slope. 
    }
    \label{fig:comp_if}
\end{figure}

We first examine Si quantum wells with smooth interfaces. At each interface, the Ge content $c_\mathrm{Ge}$ drops linearly from the barrier ($c_\mathrm{Ge}\approx 30$\%) to the well ($c_\mathrm{Ge}=0$) over the interface width $w_\mathrm{int}$. The limit $w_\mathrm{int}=0$ corresponds to the sharp interface. The Ge concentration profile for $w_\mathrm{int}=4$\,MLs is shown in Fig.~\ref{fig:model}a as an illustration. The vertical electric field is set to $F_z=8.7$\,mV/nm so that the electron probes the bottom interface. Given the cost of DFT calculations, we have computed only ten different realizations (random samplings of Ge atoms in the layers) for each interface width. The distributions of valley splittings are plotted as a function of $w_\mathrm{int}$ in Fig.~\ref{fig:comp_if}a, with dashed lines connecting the DFT and TB data for the same \YMN{atomic positions (computed with DFT)}. The correlations between the DFT and TB valley splittings are further highlighted in Fig.~\ref{fig:comp_if}b.

\begin{figure}[tbp]
    \centerline{\includegraphics[width=\linewidth]{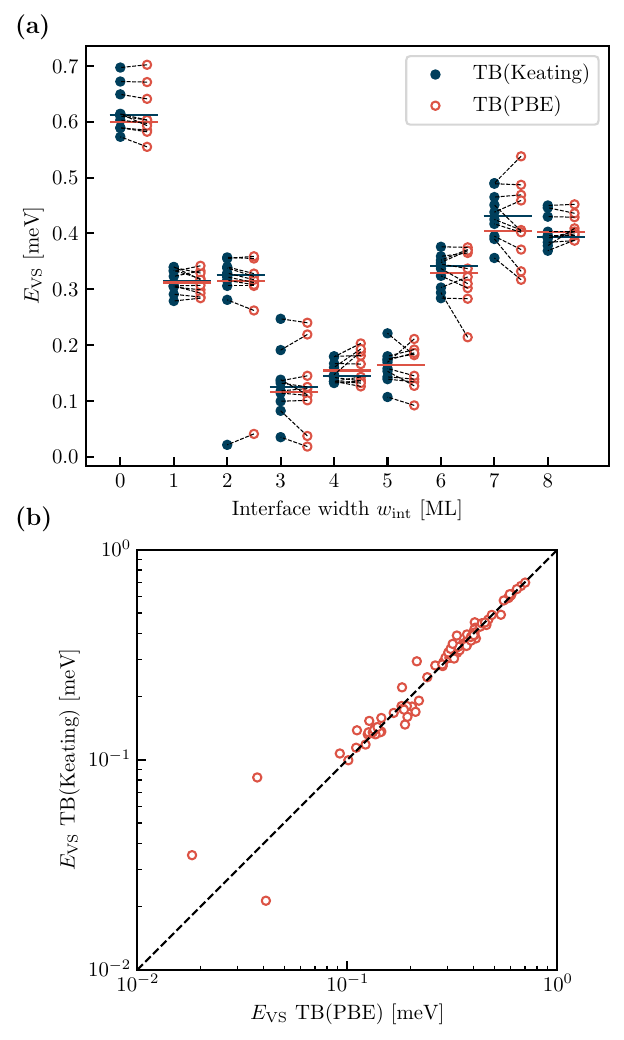}}
    \caption{
        \textbf{(a)} \YMN{TB valley splittings calculated for the same smooth interfaces as in Fig.~\ref{fig:comp_if}, starting from either Keating's VFF or PBE atomic positions. The valley splittings of the same structures are connected by dashed lines. The median value of each distribution is indicated by a horizontal bar.}
        \textbf{(b)} \YMN{Correlation between the TB valley splittings computed with Keating's VFF and DFT geometries. The dashed line is a guide to the eye with unity slope.}
    }
    \label{fig:comp_if_Keating}
\end{figure}

Several key features are illustrated by these figures. First, the median valley splitting increases for sharp interfaces and shows a dip for slightly smoothed ones ($w_\mathrm{int}=3-5$\,MLs) in line with previous studies~\cite{Lima2023,Losert2023}. Indeed, the potential of sharp interfaces exhibits larger Fourier components at the wave number $q=2k_0$ coupling the opposite $Z$-valleys [see Eq.~\eqref{eq:matrix_element}]. Yet the number of Ge atoms probed by the electron (the overlap of the squared envelope $|\Phi|^2$ with the Ge-rich MLs) is larger for graded interfaces. This gives rise to the bounce of the median valley splitting for $w_\mathrm{int}\gtrsim 5$\,MLs. These trends are in line with the expectations of the $2k_0$ theory~\cite{Losert2023}. 

\YMN{Second, DFT and TB are in reasonable quantitative agreement. The median valley splittings are, in particular, comparable in the two approaches. However, the width of the distributions are slightly larger in DFT than in TB. Moreover, there is no systematic correspondence (for a given $w_\mathrm{int}$) between the TB and DFT structures with smallest or largest valley splittings.}

\YMN{As the atomic positions used in DFT and TB are the same, these discrepancies must result from the effects of atomistic disorder on the electronic structure.} The impact of alloy disorder on the valley splittings is actually twofold. In the simplest EM approximation, alloy disorder is modeled as a potential $\delta V_\mathrm{qw}(\vec{r})$ proportional to the local deviation of the Ge fraction with respect to the target concentration (thus implicitly assuming the potential of a Ge atom is short-range and independent on its environment). To first-order in perturbation (the $2k_0$ theory), the variability of the valley splittings then results from the fluctuations of the total number of Ge atoms effectively probed by the electron in each ML. \YMN{This mechanism has been extensively discussed in quantum dots by Ref.~\cite{Losert2023}. However, the number of Ge atoms in each ML is the same whatever the realization of a given Ge concentration profile in the present design of the supercells. As the electron is not confined laterally, it thus equally probes the same number of Ge atoms for all realizations of this profile. Therefore, the dispersion of the TB and DFT valley splittings shown in Fig.~\ref{fig:comp_if} can not be explained by this counting argument. It must result from non-local corrections to the potential depending on the specific arrangement of the atoms in each ML (the behavior of a Ge atom being obviously dependent on the nature of its neighbors). The hybrid functional used in the present DFT calculations is actually non-local by nature; but more generally, DFT, as a self-consistent method, is able to account for the long range effects of the charge transfers between atoms on the potential in the structure. On the opposite, the semi-empirical TB model may capture the effects of strains but accounts explicitly for first nearest-neighbor physics only (yet not even self-consistently). The non-local corrections are expected to be more sensible in highly inhomogeneous systems such as strongly disordered alloys and interfaces, and are particularly relevant for valley physics that probe short (atomic scale) wavelengths. Although they increase the spread of the DFT valley splittings, Fig.~\ref{fig:comp_if} suggests that the effects of these corrections average to $\approx 0$ over large ensembles of alloy realizations as the DFT and TB medians are comparable for all $w_\mathrm{int}$'s.}

\YMN{The standard deviation of the valley splitting distribution is expected to scale as $1/\sqrt{S}$ with increasing the area $S=n^2a^2$ of the supercell, as the effects of alloy disorder average out~\cite{Losert2023}. We have verified this scaling law on TB calculations in larger supercells (see Appendix \ref{app:scaling}). As discussed above, the net variability in quantum dot devices results from the fluctuations of the number of Ge atoms probed by the wave function in each ML, and from the above non-local effects of atomistic disorder. The former counting contribution is expected to be the same in DFT, TB, and the $2k_0$ theory, while the latter non-local effects are completely missing in the $2k_0$ theory, and look rather poorly described by TB (taking DFT as the reference). In order to understand the importance of the different contributions, we have performed additional TB calculations in Appendix \ref{app:scaling} where the type of each atom in a given ML is chosen independently according to the Ge concentration $x$ in that ML. As a consequence, the total number $n_\mathrm{Ge}$ of Ge atoms in each ML is not fixed any more and follows a binomial distribution $B(2n^2, x)$ whose relative standard deviation $\sqrt{\langle\delta n_\mathrm{Ge}^2\rangle}/\langle n_\mathrm{Ge}\rangle$ also scales as $1/\sqrt{S}$. The electron thus now probes a different number of Ge atoms in each random realization of the supercell. It turns out that the variability of the valley splittings induced by these (counting) fluctuations is about one order of magnitude larger than the variability due to the non-local effects of atomistic disorder. Therefore, we conclude that, despite the poor one-to-one correspondences in Fig.~\ref{fig:comp_if}, TB remains reasonably accurate for the statistical analysis of valley splittings in quantum dots hosted in smoothed Si/SiGe quantum wells whose variability is typically dominated by Ge counting fluctuations. Moreover, we have demonstrated in Ref.~\cite{Salamone25} that the so-called 2 bands $\vec{k}\cdot\vec{p}$ model~\cite{Hensel1965,Sverdlov2007,Sverdlov2015}, which is a (yet non-perturbative) implementation of the EM $2k_0$ theory, yields valley splitting statistics in excellent agreement with TB models in quantum dots in various Si/SiGe heterostructures.}

\YMN{We also compare the TB valley splittings calculated with DFT and with Keating's VFF atomic positions in Fig.~\ref{fig:comp_if_Keating}. The bond lengths, bond stretching and bond bending constants of the VFF are taken from Ref.~\cite{Niquet2009}. The lattice parameters and elastic constants of the VFF and DFT frameworks are thus slightly different. Yet the TB valley splittings calculated from DFT and VFF positions are globally consistent (although the data are slightly more dispersed on average with DFT than with VFF positions). This demonstrates that Keating's VFF provides a reasonable starting point to compute TB valley splittings for larger systems out of reach of DFT.}

\subsection{Wiggle wells}

Quantum wells with oscillating Ge concentrations -- so-called wiggle wells -- are a promising strategy to enhance the detrimentally low valley splitting in Si/SiGe heterostructures~\cite{McJunkin2022,Feng2022,Losert2023}. As the concentration of Ge is modulated within the well, significant valley splittings can be achieved at low or even zero electric field $F_z$. 

Here we consider wiggle wells whose Ge concentrations oscillate between 0 and 10\% with wave number $q_w$ ranging from 2 to 20\,nm$^{-1}$. The well is still confined by abrupt, unmodulated Si$_{0.7}$Ge$_{0.3}$ barriers and the electric field is $F_z=0$. We discuss in Appendix~\ref{app:spectra} the actual Ge concentration profiles achieved in our small supercells and their relevance to the original wiggle well design.

\begin{figure}[tbp]
	\centerline{\includegraphics[width=\linewidth]{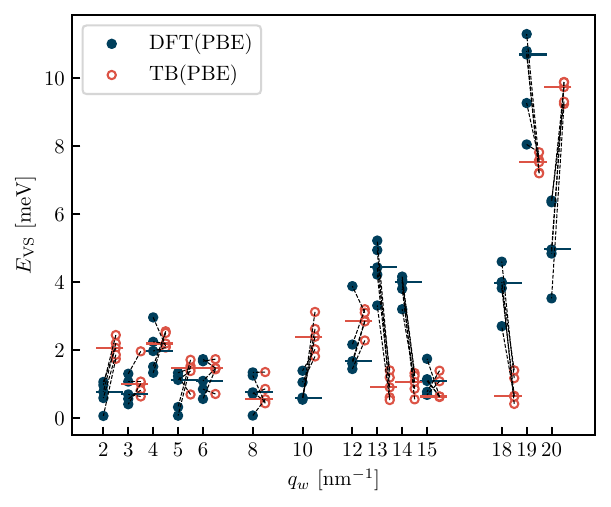}}
	\caption{
     DFT and TB valley splittings $E_\mathrm{VS}$ calculated in wiggle wells with wave number $q_w$ (see text). \YMN{The atomic positions relaxed with the PBE exchange-correlation functional are used as input for DFT and TB calculations.} The plot shows the distribution of five random realizations of each $q_w$, with the DFT and TB valley splittings of the same structures connected by dashed lines. The median value of each distribution is indicated by a horizontal bar. 
	}
	\label{fig:comp_wiggle}
\end{figure}

The valley splittings calculated for 5 random realizations of each wave number are plotted in Fig.~\ref{fig:comp_wiggle}. As for smoothed quantum wells, the same DFT and TB structures are connected by dashed lines. The correlations between DFT and TB valley splittings are shown in Fig.~\ref{fig:comp_wiggle}b. As expected, the DFT valley splittings exhibit a main peak at $q_w\approx 20$\,nm$^{-1}$ when the wiggle well modulations are resonant with the wave number $q=2k_0$ coupling opposite valleys in Eq.~\eqref{eq:matrix_element}. The DFT data also show secondary peaks at $q_w\approx 4$\,nm$^{-1}$ and $q_w\approx 12-14$\,nm$^{-1}$~\cite{McJunkin2022,Feng2022,Losert2023}. The first one results from umklapp processes (inter-valley scattering by a modulation with wave number $q=2|k_0-2\pi/a_0|$), and is enabled by alloy disorder (as it gets suppressed in a pure diamond lattice with inversion-symmetric Bloch functions). The second one is a sub-harmonic of the main peak at $q_w\approx 20$\,nm$^{-1}$. None of these two peaks are captured by the $2k_0$ theory. 

The agreement between the DFT and TB data is much worse for wiggle than for smoothed quantum wells, especially at large $q_w$. \YMN{Nevertheless, the TB model captures the main peak at $q_w\approx 19$\,nm$^{-1}$ as well as the secondary peaks. They appear shifted with respect to DFT, however. As a matter of fact,} the valley splittings of wiggle wells are ruled by the resonance conditions $q_w\approx 2k_0$, $q_w\approx k_0$ or $q_w\approx 2|k_0-2\pi/a_0|$. They thus depend on the exact value of $k_0$, because the Fourier transform of the wiggle well profile exhibits a sharp peak at $q=q_w$ \YMN{(see Appendix \ref{app:spectra})} and does not vary continuously with $q_w$ as the Ge content is discretized in steps of 5.56\% in the present simulations. The position of the conduction band minima is, however, not exactly the same in DFT and TB ($2k_0\approx 19.4$\,nm$^{-1}$ in TB and $2k_0\approx 20.1$\,nm$^{-1}$ in DFT as inferred from the period of the valley oscillations in Fig.~\ref{fig:DFTsolutions}). This makes a quantitative comparison between DFT and TB  more difficult than for smoothed quantum wells with a broader Fourier spectrum. 

Moreover, strong valley-orbit hybridization effects (discussed in the next paragraph) reshape the main peak around $q_w\approx 18-20$\,nm$^{-1}$ (both in DFT and TB), as the valley states of the ground and excited subbands repel each other and mix. Therefore, valley(-orbit) splittings in very rapidly varying potentials are (not unexpectedly) much more dependent on atomistic and band structure details (value of $k_0$, subband splittings, ...). While TB remains suitable for a semi-quantitative estimate of valley splittings in such potentials, a fully quantitative assessment (with respect to DFT -- which has its own limitations \footnote{Although DFT can account for the interplay between structural and electronic properties from first principles (thus in a much more versatile way than TB), it does not exactly reproduce, as discussed in Section \ref{sec:interpretation}, lattice parameters or band offsets. The results may, moreover, depend on the choice of exchange-correlation functional.} -- or experiment) thus appears more challenging.

\subsection{Valley-orbit mixing}

\begin{figure}[]
    \centerline{\includegraphics[width=\linewidth]{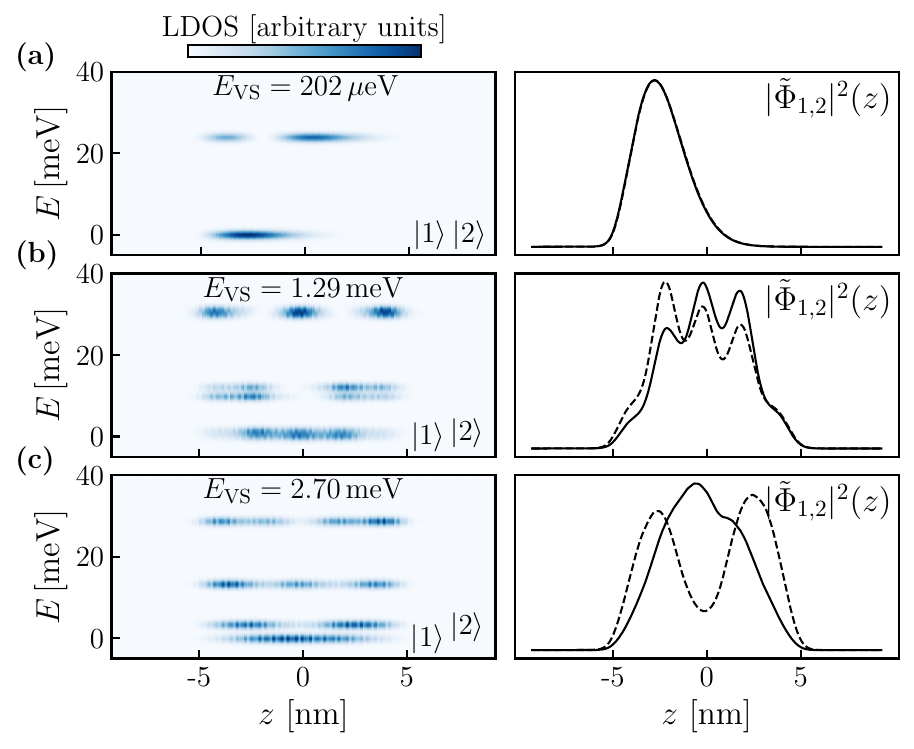}}
    \caption{
    DFT LDOS (left) and squared envelopes $|\tilde\Phi_{1,2}|^2$ of the valley-orbit ground-states (solid and dashed lines on the right) for three different quantum wells with various designs and electric fields.
    \textbf{(a)} Si well with smooth interfaces ($w_\mathrm{int}=3$\,MLs) at $F_z=8.76$\,mV/nm. The excited sub-bands are separated by about 24\,meV. The DFT envelopes overlap almost perfectly. 
    \textbf{(b)} Wiggle well with wave number $q_w=3$\,nm$^{-1}$ at zero electric field. The valley splitting increases ($E_\mathrm{VS}=1.29$\,meV) but the orbital splitting is smaller (10\,meV) because of the low electric field. The envelopes $\tilde\Phi_{1,2}$ are modulated by the oscillating Ge concentration and show significant differences due to valley-orbit mixing.
    \textbf{(c)} Wiggle well with wave number $q_w\approx 2k_0=18$nm$^{-1}$ at $F_z=0$. Valley-orbit mixing is pervasive at all energies.}
    \label{valley-orbit}
\end{figure}

\begin{figure}[]
    \centerline{\includegraphics[width=\linewidth]{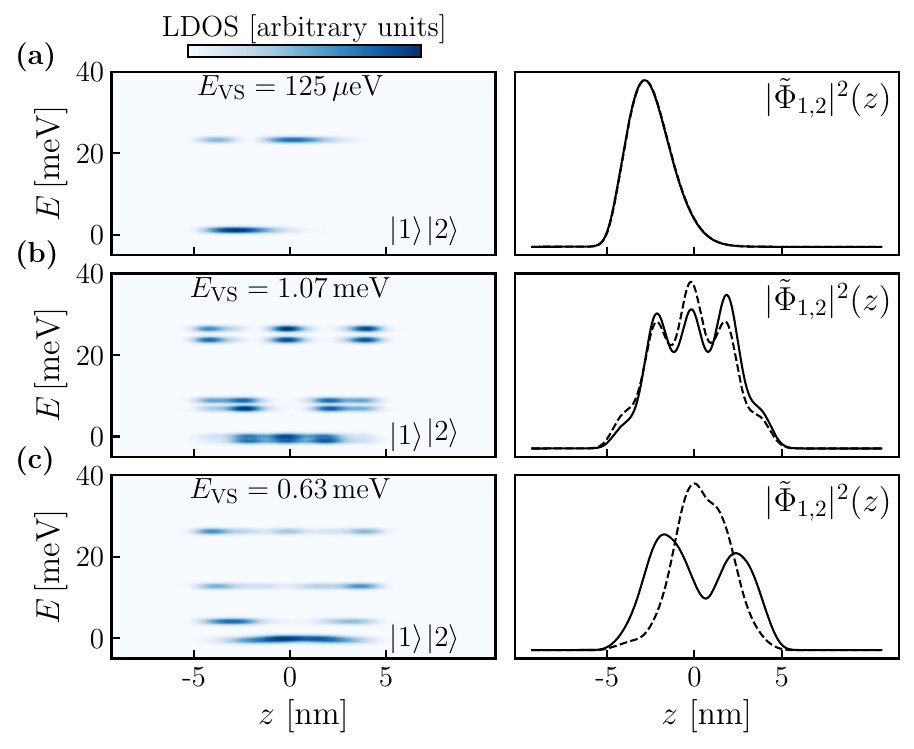}}
    \caption{
    TB LDOS (left) and squared envelopes $|\tilde\Phi_{1,2}|^2$ of the valley-orbit ground-states (solid and dashed lines on the right) for the same structures as in Fig.~\ref{valley-orbit}.
    \textbf{(a)} Si well with smooth interfaces ($w_\mathrm{int}=3$\,MLs) at $F_z=8.76$\,mV/nm. The excited sub-bands are separated by about 22\,meV, similar to DFT. 
    \textbf{(b)} Wiggle well with wave number $q_w=3$\,nm$^{-1}$ at zero electric field. The valley splitting  ($E_\mathrm{VS}=1.07$\,meV) as well as the envelopes are similar to DFT.
    \textbf{(c)} Wiggle well with wave number $q_w\approx 2k_0=18$nm$^{-1}$ at $F_z=0$. Note that the valley splitting is smaller than in DFT, but the order of the two valley states is different.}
    \label{valley-orbit_TB}
\end{figure}

A more detailed analysis of the DFT data for configurations with large valley splitting reveals the signatures of strong valley-orbit mixing, that is the admixture of higher excited orbitals resulting in valley states with different envelopes~\cite{PhysRevB.81.115324,PhysRevB.88.035310,Tariq2022}. To illustrate the effects of valley-orbit mixing, we compare three specific configurations. First, a Si well with smooth interfaces ($w_\mathrm{int}=3$\,MLs) at electric field $F_z=8.76$\,mV/nm, whose envelope functions and LDOS are plotted in Fig.~\ref{valley-orbit}a. Such structures give rise to much weaker inter-valley scattering than wiggle wells ($E_\mathrm{VS}=202\,\mu$eV, not resolved on the LDOS), while the strong electric field splits the lowest orbital sub-bands by $>20$ meV. The envelopes $\tilde\Phi_{1,2}$ of the two valley ground-states are barely distinguishable. Next, we consider a wiggle well with wave number $q_w=3$\,nm$^{-1}$ at zero electric field (Fig.~\ref{valley-orbit}b). At this field, the sub-band splitting is much lower ($10$\,meV), and the valley splitting is enhanced by the wiggle well design ($E_\mathrm{VS}=1.29$\,meV)~\cite{Losert2023}. This allows for a small but differentiated admixture of higher-lying sub-bands by inter-valley scattering, resulting in slightly different envelopes $\tilde\Phi_{1,2}$. Finally, we investigate a wiggle well with large wave number $q_w=18$\,nm$^{-1}$ (Fig.~\ref{valley-orbit}c). The LDOS now displays the fingerprints of valley-orbit mixing at all energies, with the valley states of the ground and excited subbands anti-crossing each other and mixing \footnote{Note that we still define in that case the ``valley splitting'' as the difference between the lowest two eigenstates, since valley and orbital degrees of freedom may get significantly mixed, and since this is the experimentally relevant quantity for spin qubits.}.

\YMN{The effects of valley-orbit mixing are also, to a large extent, captured by the TB model (see Fig.~\ref{valley-orbit_TB}), but are, by design, missing in the perturbative $2k_0$ theory (which assumes the same envelopes for both valley states). In Figs.~\ref{valley-orbit_TB}a,b, the TB LDOS is in good overall agreement with DFT, but in Fig.~\ref{valley-orbit_TB}c (wiggle well with fast oscillations at $q_w=18$\,nm$^{-1}$), the TB LDOS is significantly different from DFT over the whole energy range. In particular, the lowest two states have crossed and changed order (but their small splitting can not be resolved on Fig.~\ref{valley-orbit_TB}c). As discussed above, the quantitative agreement between TB and DFT degrades with increasing $q_w$, although both continue to describe the same physics.} The valley-orbit mixing tends to strengthen when the ratio between the valley and orbital splitting increases. Interestingly, valley-orbit mixing is usually suppressed at high electric fields because the orbital gaps open faster than the valley splittings when increasing $F_z$.

In quantum dots, the (minimal) orbital splitting is, however, set by lateral rather than by vertical confinement and is usually as low as $\approx 1-2$\,meV. If the confinement potential is fully separable, each envelope can be split into the product of an in-plane component $\Phi_\parallel(x,y)$ and an out-of-plane component $\Phi_\perp(z)$. Inter-valley scattering does not mix $\Phi_\parallel$ envelopes and the relevant orbital splitting remains the sub-band splitting between the $\Phi_\perp(z)$'s. However, realistic confinement potentials are not separable (as the electrostatics of the devices is not, and because of charge and alloy disorder)~\cite{martinez2022hole}, so that the vertical and in-plane motions can be significantly coupled. Inter-valley scattering may then mix the nearby (mostly in-plane) excitations. The prevalence of such valley-orbit mixing effects in quantum dots has been recently demonstrated in TB and two bands $\vec{k}\cdot\vec{p}$ calculations~\cite{Salamone25}.

\section{Conclusions}

We have calculated the valley splittings of planar SiGe heterostructures with DFT using a hybrid exchange-correlation functional. We have compared the DFT splittings with semi-empirical TB calculations. We find that both atomistic methods reproduce the same global trends in line with the expectations of the ``$2k_0$'' EM theory. \YMN{However, the agreement between DFT and TB is strongly dependent on the Ge concentration profile. It is much better for profiles with broad Fourier transforms (such as smoothed interfaces) than for rapidly modulated Ge concentration profiles with peaked Fourier transforms (such as wiggle wells with large wave numbers). These discrepancies are mostly a consequence of the different microscopic descriptions of alloys. In particular, DFT can better account for charge transfers and for long-range effects than semi-empirical TB (not to mention the EM). Nevertheless, TB (and the $2k_0$ theory) can provide a reasonable basis for the statistical analysis of valley splittings in quantum dots hosted in a wide range of Si/SiGe heterostructures. DFT and TB also highlight the effects of strong valley-orbit mixing once the inter-valley scattering strength is large enough with respect to the orbital splittings. Although such valley-orbit mixing effects are not captured by the original (perturbative) $2k_0$ theory, they can be accounted for in non-perturbative implementations such as the two bands $\vec{k}\cdot\vec{p}$ model discussed in Ref.~\cite{Salamone25}.}

\section{Acknowledgments}

We thank Dominic Waldh\"or for helpful discussions. This project has received funding from the European Research Council (ERC) under grant agreement no. 101055379, and from the French National Research Agency under the program ``France 2030'' (PEPR PRESQUILE -- ANR-22-PETQ-0002). Calculations were performed using supercomputer resources provided by the Vienna Scientific Cluster (VSC). L.\,C. gratefully acknowledges support from Institut français d'Autriche.

\appendix

\section{The vertical electric field within DFT}
\label{app:field}
\begin{figure}[ht]
    \centerline{\includegraphics[width=\linewidth]{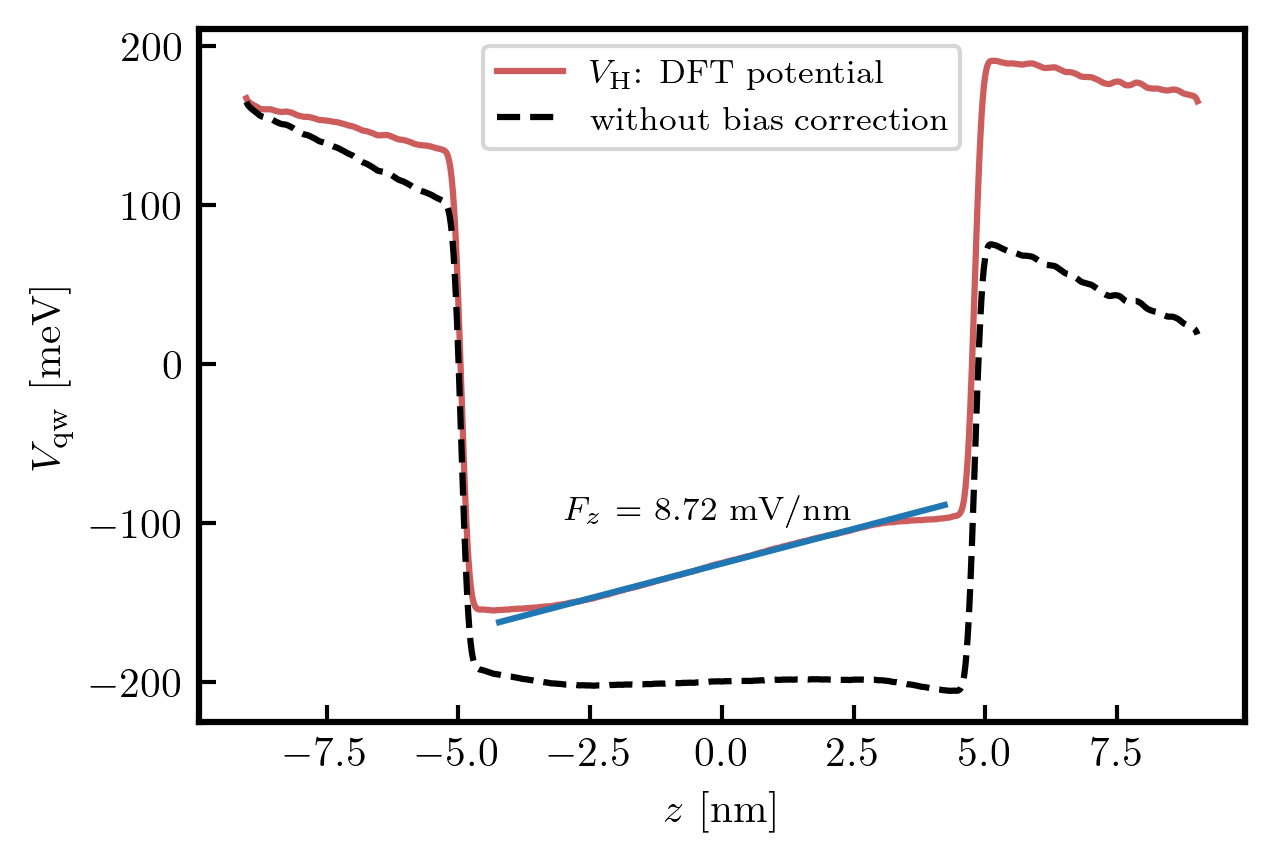}}
    \caption{
        Electrostatic (Hartree) potential profiles across a biased SiGe/Si/SiGe superlattice with sharp interfaces, with (red line) and without (black dashed line) bias correction. The potential is averaged in the cross-section of the superlattice and convoluted with a gaussian to wash out atomic-scale variations (see Section \ref{sec:methodology}). The black dashed line depicts the expected distribution of electric field in the stacked materials according to their thickness and permittivity. Such a potential however jumps at the edges of the periodic supercell. The bias correction implemented in the DFT code restores the continuity of the electrostatic potential across the edges of the supercell. The blue line is a fit $F_z=8.72$\,mV/nm to the effective electric field in the Si well.
    }
    \label{fig:DFT_potential}
\end{figure}

CP2K~\cite{CP2K} uses the Berry-phase formalism~\cite{Vanderbilt_solids_1993, Vanderbilt_surface-charge_1993} to enable the application of a voltage bias across the whole simulation cell~\cite{Souza2002, Umari2002} and circumvent the problems arising from potential jumps in periodic systems \footnote{
	\label{footnote_continuous}
	A constant bias, i.e.~a constant displacement field across the whole simulation cell is only consistent with periodic boundary conditions if the electrostatic potential is discontinuous. Such discontinuities lead to unphysical results.
}. The implementation in CP2K is based on imposing closed-circuit boundary conditions with a constant bias. This bias polarizes the dielectric layers and thereby leads to the accumulation of charge at the interfaces. The potential thus drops according to each layer thickness and relative permittivity. However, by correcting for the applied bias, the potential remains continuous at the edges of the supercell, as illustrated in Fig.~\ref{fig:DFT_potential}. With this prescription, the effective electric field in the DFT calculation (as defined by the gradient of the Hartree potential $V_\mathrm{H}$) shows opposite orientations in the Si and SiGe layer.

\begin{figure}[ht]
    \centerline{\includegraphics[width=\linewidth]{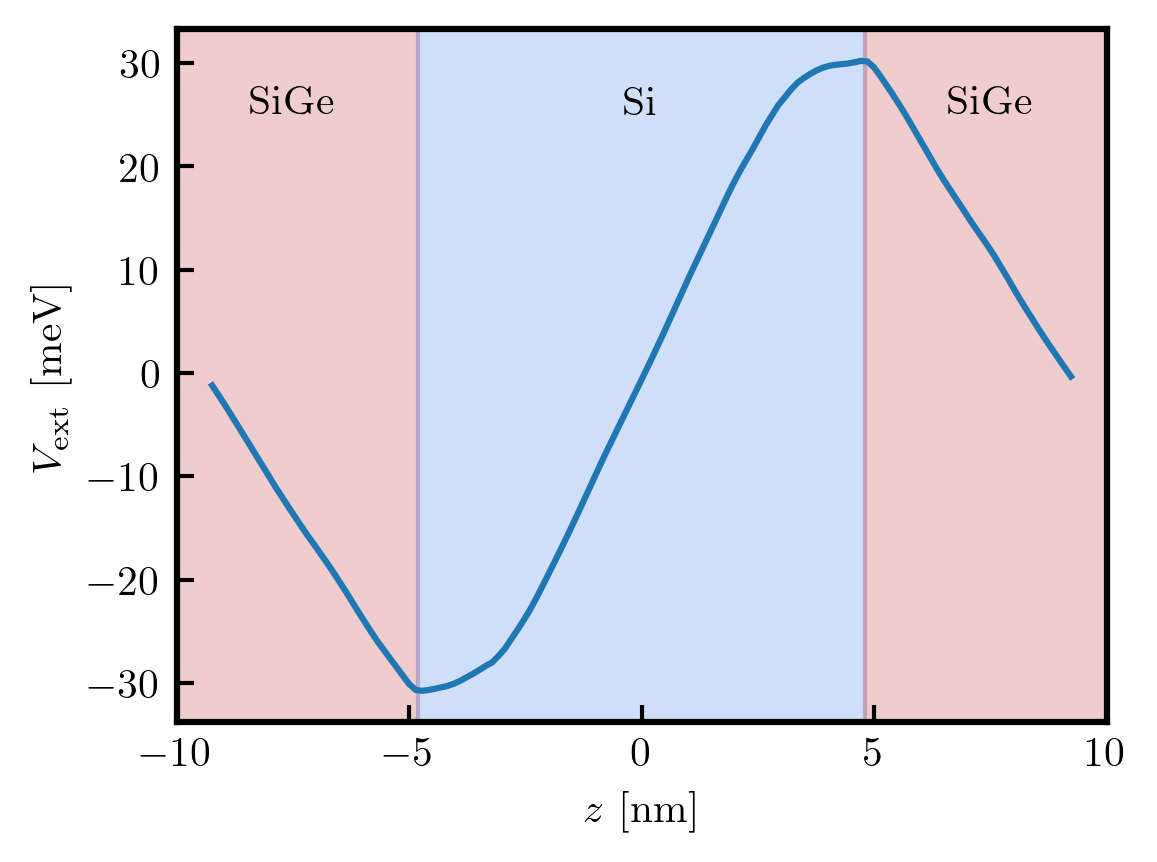}}
    \caption{
    External potential profile $V_\mathrm{ext}$ extracted in the same SiGe/Si/SiGe superlattice as in Fig.~\ref{fig:DFT_potential}. The bias corrected potential profile is continuous and periodic. As a consequence, the electric field has opposite orientations in Si and SiGe. This sawtooth potential is used as input in the TB calculations for consistent comparisons.
    }
    \label{fig:electric_field_profile}
\end{figure}

Because the Hartree potential gathers contributions from both the quantum well confinement and the external bias, the pure external potential profile $V_\mathrm{ext}$ is obtained by subtracting the quantum well potential $V_\mathrm{H,0}$ at zero electric field, as shown in Fig.~\ref{fig:electric_field_profile}. The electric fields quoted in this work are the average slope of $V_\mathrm{ext}$ in the Si well (blue line on Fig.~\ref{fig:DFT_potential}). For consistent comparisons, the sawtooth potential $V_\mathrm{ext}$ extracted from the DFT calculation is transferred to the corresponding TB Hamiltonian as a shift of the orbital energies of the atoms.

\section{Scaling of the TB data with respect to the supercell size}
\label{app:scaling}

\begin{figure}[ht]
    \centerline{\includegraphics[width=.9\linewidth]{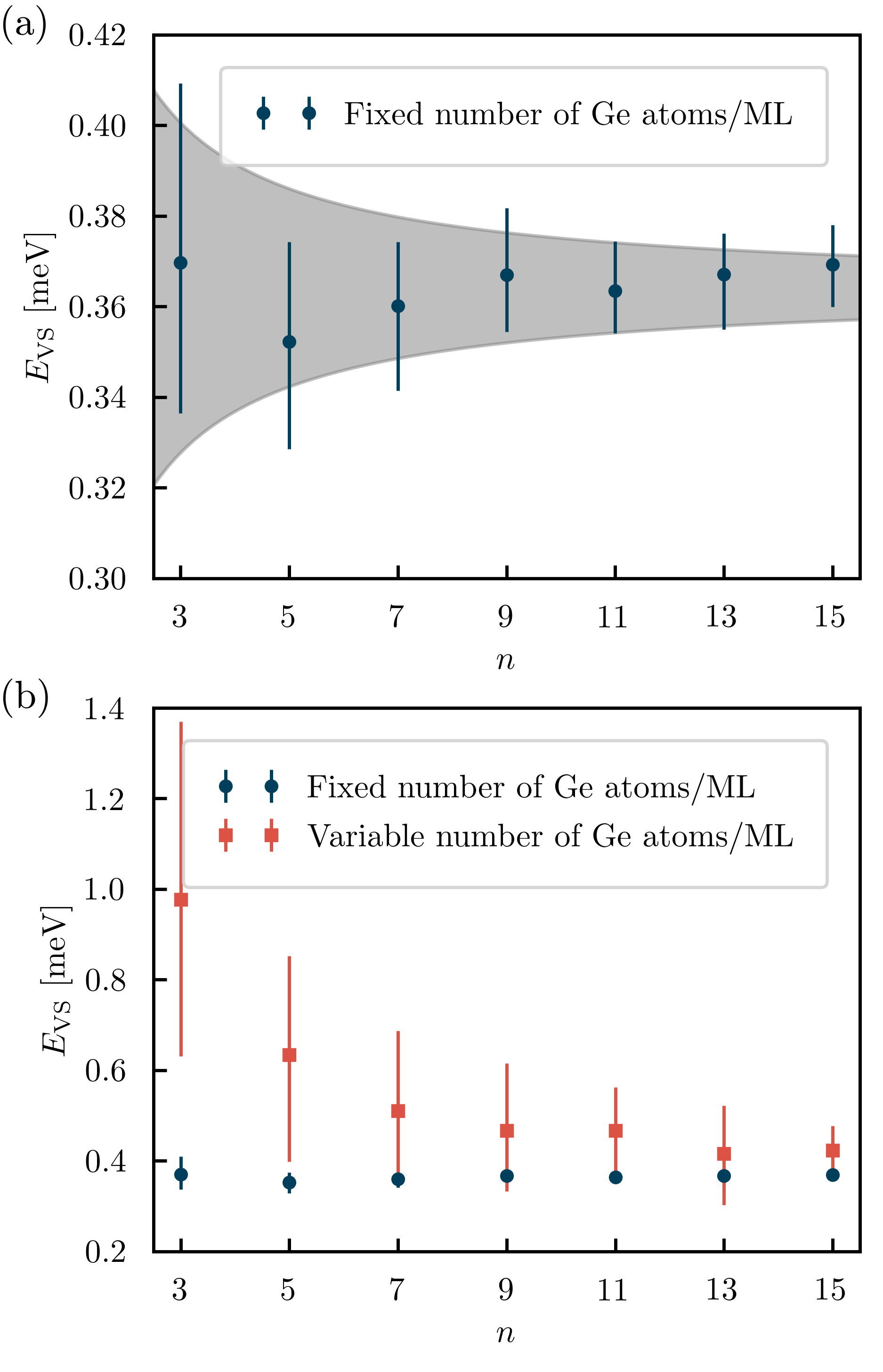}}
    \caption{\textbf{(a)} Median (dots) and inter-quartile range (error bars) of the TB valley splittings computed in smoothed wells with interface width $w_\mathrm{int}=6$\,ML, as a function of the side $n$ of the supercell. The electric field is $E_z=8.7$\,mV/nm. The structures were relaxed with Keating's valence force field and the statistics computed over 128 realizations of the alloy. The shaded gray area outlines the expected $\propto 1/n$ dependence of the inter-quartile range. The number of Ge atoms per ML is fixed and set by the Ge concentration profile. \textbf{(b)} Comparison between the statistics of panel (a) (blue) and the statistics computed assuming a binomial distribution of the number of Ge atoms per ML (red).
    }
    \label{fig:scaling}
\end{figure}

\YMN{In this appendix, we discuss the dependence of the statistics of the valley splittings on the side $n$ of the supercell. As a representative illustration, we plot in Fig.~\ref{fig:scaling}a the median and inter-quartile range of the TB valley splittings computed in smoothed wells with interface width $w_\mathrm{int}=6$\,MLs. These structures, that are out of reach of DFT for $n>3$, were relaxed with Keating's valence force field and the statistics computed over 128 realizations of the alloy.}

\YMN{The median valley splitting appears already well converged in the smallest $3a\times 3a$ supercells considered in the main text. As expected from statistical arguments, the inter-quartile range, that quantifies the width of the valley splitting distribution, decreases as $1/n$ (or equivalently as $1/\sqrt{S}$, with $S=n^2a^2$ the surface of the supercell). A similar scaling was obtained for wiggle wells in Ref.~\cite{Gradwohl2024}. In these simulations, the number of Ge atoms in a given ML is fixed (and defined by the Ge concentration); only the distribution of these atoms in the ML thus changes from sample to sample. As discussed in the main text, Fig.~\ref{fig:scaling}a thus highlights the non-local effects of atomistic disorder.}

\YMN{In a quantum dot with finite area $S$, the number of Ge atoms probed by the electron in a given ML however fluctuates from sample to sample, and follows a binomial distribution $B(2S/a^2, x)$ where $x$ is the Ge concentration in that ML. In order to emulate this behavior in the present 2D calculations, we draw independently the nature of each atom of a ML according to $x$. The statistics of the TB valley splittings in such structures are plotted in Fig.~\ref{fig:scaling}b. The spread of the data is almost an order of magnitude larger than in Fig.~\ref{fig:scaling}a; more importantly, the median valley splitting is much larger in small supercells. Indeed, the matrix element $\Delta$ in Eq.~(\ref{eq:matrix_element}) is a complex number with an approximately gaussian distribution in the complex plane; the valley splitting $E_\mathrm{VS}=2|\Delta|$ thus follows a Rician distribution whose average $2\langle|\Delta|\rangle$ is expected to be much larger than $|\langle\Delta\rangle|$ if the standard deviation $\sigma_{|\Delta|}$ is large~\cite{Losert2023}. Hence large variabilities limit reproducibility but can result in a (beneficial) increase of the average valley splitting of weakly coupled valleys. The (Ge counting) fluctuations also decrease as $1/n$, and the median valley splitting asymptotically converges to the same limit as in Fig.~\ref{fig:scaling}a. Given the fact that the variability is usually limited in quantum dots by these counting fluctuations (that are the same in DFT, TB and the EM), we conclude that TB (and the $2k_0$ theory) provide a reasonable account of the statistics of valley splittings.}


\section{Wiggle well potentials}
\label{app:spectra}
\begin{figure}[htb]
    \centerline{\includegraphics[width=\linewidth]{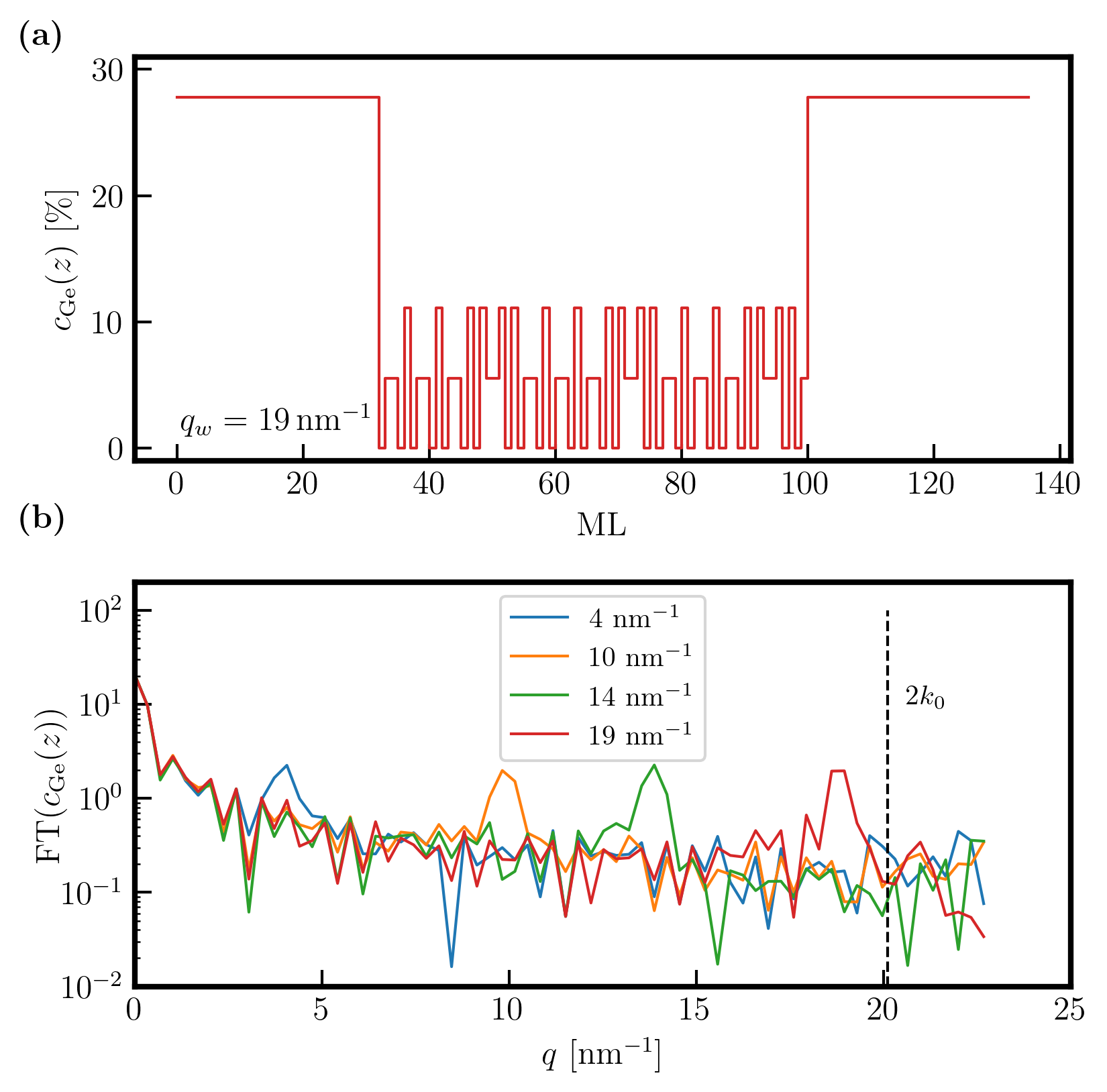}}
    \caption{
    \textbf{(a)} A discretized wiggle well Ge profile across the simulation cell ($q_w=19$\,nm$^{-1}$, $\Delta_\mathrm{Ge}=10$\%). 
    \textbf{(b)} Discrete Fourier transform of some representative wiggle well profiles with different wave numbers $q_w$. Despite the small supercell size which allows only rough steps of 5.56\% in the Ge content, each structure shows a pronounced peak at the expected $q=q_w$. 
    }
    \label{fig:spectra}
\end{figure}

Wiggle wells with sinusoidal Ge concentrations
\begin{equation}
c_\mathrm{Ge}(z)=\frac{\Delta_\mathrm{Ge}}{2}\left(1-\cos q_w z\right)
\end{equation}
can not be represented exactly in our small DFT supercells, which allow only for step-like Ge distributions with increments $\delta c_\mathrm{Ge}=1/18\approx5.56$\%. The actual concentration profile implemented for $\Delta_\mathrm{Ge}=10$\% and $q_w=19$\,nm$^{-1}$ is plotted in Fig.~\ref{fig:spectra}a as an example. For all wiggle wells discussed in this work, $z=0$ corresponds to the bottom interface. Despite strong discretization, the Fourier transform of this profile exhibits the expected peak at wave number $q=q_w$ (see Fig.~\ref{fig:spectra}b). Therefore, small supercells also give rise to enhanced inter-valley scattering when $q_w\approx 2k_0$. The TB and DFT calculations are performed on the same structural models (same Ge distributions and atomic positions) for unbiased comparisons.

\bibliography{my.bib}

\end{document}